%% file: sample-base.tex
\documentclass[sigconf,natbib=true,anonymous=false]{acmart}
\AtBeginDocument{%
  \providecommand\BibTeX{{%
    \normalfont B\kern-0.5em{\scshape i\kern-0.25em b}\kern-0.8em\TeX}}}

\setcopyright{acmcopyright}
\copyrightyear{2018}
\acmYear{2018}
\acmDOI{XXXXXXX.XXXXXXX}

\acmConference[Conference acronym 'XX]{Make sure to enter the correct
  conference title from your rights confirmation emai}{June 03--05,
  2018}{Woodstock, NY}
\acmBooktitle{Woodstock '18: ACM Symposium on Neural Gaze Detection,
 June 03--05, 2018, Woodstock, NY} 
\acmPrice{15.00}
\acmISBN{978-1-4503-XXXX-X/18/06}

\usepackage[utf8]{inputenc} % allow utf-8 input
\usepackage[T1]{fontenc}    % use 8-bit T1 fonts
\usepackage{hyperref}       % hyperlinks
\usepackage{url}            % simple URL typesetting
\usepackage{booktabs}       % professional-quality tables
\usepackage{amsfonts}       % blackboard math symbols
\usepackage{nicefrac}       % compact symbols for 1/2, etc.
\usepackage{microtype}      % microtypography
\usepackage{xcolor}         % colors
\usepackage{graphicx}
\usepackage{amsthm}
\usepackage{amsmath}
\usepackage{subfigure}
\usepackage{cleveref}
\usepackage{caption}
\usepackage{threeparttable}
\usepackage{comment}
\usepackage{multirow}
\usepackage{booktabs}
\usepackage{enumitem}
\usepackage{wrapfig}

\usepackage{adjustbox}
\usepackage{bbding}

\newtheorem{definition}{Definition}
\newcommand{\indep}{\perp \!\!\! \perp}
\newcommand{\nind}{\centernot{\indep}}
\newcommand{\ly}{}

\crefname{equation}{eqn.}{Eqn.}

\graphicspath{{image/}}
\begin{document}

\title{A Semi-Synthetic Dataset Generation Framework for Causal Inference in Recommender Systems}

\author{Yan Lyu}
\authornote{Both authors contributed equally to this research.}
\affiliation{%
  \institution{Peking University}
  \country{China}
}
\author{Sunhao Dai}
\authornotemark[1]
\affiliation{%
  \institution{Renmin University of China}
  \country{China}
}

\author{Peng Wu}
\affiliation{%
  \institution{Peking University}
  \country{China}
}

\author{Quanyu Dai}
\affiliation{%
  \institution{Huawei Noah’s Ark Lab}
  \country{China}
}

\author{Yuhao Deng}
\affiliation{%
  \institution{Peking University}
  \country{China}
}

\author{Wenjie Hu}
\affiliation{%
  \institution{Peking University}
  \country{China}
}

\author{Zhenhua Dong}
\affiliation{%
  \institution{Huawei Noah’s Ark Lab}
  \country{China}
}

\author{Jun Xu}
\affiliation{%
  \institution{Renmin University of China}
  \country{China}
}

\author{Shengyu Zhu}
\affiliation{%
  \institution{Huawei Noah’s Ark Lab}
  \country{China}
}

\author{Xiao-hua Zhou}
\affiliation{%
  \institution{Peking University}
  \country{China}
}
% \thanks{$\dagger$ Corresponding author: Shengyu Zhu and Xiao-hua Zhou. Email: zhushengyu@huawei.com and azhou@math.pku.edu.cn}
\renewcommand{\shortauthors}{Lyu and Dai, et al.}

\begin{abstract}
  Accurate recommendation and reliable explanation are two key issues for modern recommender systems. However, most recommendation benchmarks only concern the prediction of user-item ratings while omitting the underlying causes behind the ratings. For example, the widely-used Yahoo!R3 dataset contains little information on the causes of the user-movie ratings. A solution could be to conduct surveys and require the users to provide such information. In practice, the user surveys can hardly avoid compliance issues and  sparse user responses, which greatly hinders the exploration of causality-based recommendation. To better support the studies of causal inference and further explanations in recommender systems, we  propose a novel semi-synthetic data generation framework for recommender systems where causal graphical models with missingness are employed to describe the causal mechanism of practical recommendation scenarios. To illustrate the use of our framework, we construct a semi-synthetic dataset with Causal Tags And Ratings (CTAR), based on the movies as well as their descriptive tags and rating information collected from a famous movie rating website. Using the collected data and the causal graph, the user-item-ratings and their corresponding user-item-tags are automatically generated, which provides the reasons (selected tags) why the user rates the items. Descriptive statistics and baseline results regarding the CTAR dataset are also reported. The proposed data generation framework is not limited to recommendation, and the released APIs can be used to generate customized datasets for other research tasks.
\end{abstract}

% <concept_id>10010147.10010178</concept_id>
% <concept_desc>Computing methodologies~Artificial intelligence</concept_desc>

% \ccsdesc[500]{Computing methodologies~Artificial intelligence}

\keywords{causal inference, missingness graph, recommender systems, semi-synthetic dataset generation}


\maketitle

\section{Introduction}\label{Introduction}
Recommendation models serve as a core component in modern recommender systems. Most of these models, either factorization machine \cite{rendle2010factorization,SIGIR_RendleGFS11,RecSys_JuanZCL16} or neural network based models \cite{cheng2016wide,guo2018deepfm,RecSys_HuangZZ19},  target at user-item rating prediction tasks, e.g., whether a user would give high ratings to some movies. Although they have achieved impressive results in both benchmarked datasets and real products, they often omit the underlying \emph{causes} behind users' ratings and generally lack reliable explanations. The underlying causes, like ``the user is particularly interested in romantic movies'', can be very 
useful in achieving both accurate and explainable recommendations. However, even when we are concerned with the question why he/she likes the item and proceeds to develop methods to infer the reason, existing datasets such as Yahoo!R3 \cite{marlin2009collaborative} and Coat \cite{schnabel2016recommendations} cannot provide the true causes to evaluate these methods and hence are less than useful to this purpose.

A tentative solution might be to conduct a survey where the subjects are asked to explicitly input the causes that make he/she give a high rating. As we will discuss in \Cref{datasetdesign}, in practice the obtained survey data are likely to be sparse and fail to serve as a statistically reliable benchmark. There also exist possible compliance issues, e.g., some users may give arbitrary ratings despite that they are expected to provide their true preferences, which may affect the purpose of evaluation.

\sloppy
% In this work, we propose a semi-synthetic dataset with Causal Tags And Ratings (CTAR) that aims to  better support causal inference and further explanations in recommender systems. For example, we may consider to infer the cause of user preference through estimating the causal effect of each tag, or using the \emph{do}-operator terminology \cite{pearl2009}: we are interested in the \zhushengyu{name for this?} \[\mathbb E[\text{Rating}|do(\text{tag existing in the movie}), \text{user}] - \mathbb E[\text{Rating}|do(\text{tag absent in the movie}),\text{user}]\]. Our dataset is generated from a causal perspective with guaranteed causal interpretations. We first collect observational data containing movies with their descriptive tags and rating information from a famous movie rating website, where users can rate and apply tags to the movies. We then use causal graphical models with missing mechanism to mimic the data generation procedure in practical scenarios, with hyper-parameters determined according to the collected movie and tag data. Causal graphical models work in a disentangled way and provide a flexible framework for synthetic dataset generation---we can simply add corresponding nodes and edges to the graphical model if we would like to include more kinds of biases in the CTAR dataset. Moreover, graphical models can help verify causality assumptions, e.g., we can easily verify conditional independencies using \emph{d}-separation criterion \cite{pearl2009}.
In this work, we propose a novel semi-synthetic data generation framework that aims to  better support causal inference and further explanations in recommender systems. For example, we may consider to infer the cause of user preference by estimating the causal effect of each tag; using the \emph{do}-operator terminology \cite{pearl2009}, we are interested in $\mathbb E[Rating\mid user~u,~do(with~tag~T)] - \mathbb E[Rating\mid user~u,~do(without~tag~T)]$. Our framework includes causal graphical models with missing mechanism to mimic the data generation procedure in practical scenarios. Causal graphical models work in a disentangled way and provide a flexible framework for synthetic dataset generation---we can simply add corresponding nodes and edges to the graphical model if we would like to include more kinds of biases in the framework. Moreover, graphical models can help verify causality assumptions, e.g., we can easily verify conditional independencies using \emph{d}-separation criterion \cite{pearl2009}. To  illustrate the usage of our framework, a semi-synthetic dataset with Causal Tags And Ratings (CTAR) is constructed based on our framework. In particular, we first collect observational data containing movies with descriptive tags and rating information from a famous movie rating website where users can rate and apply tags to the movies, and then use the collected movie and tag data to determine the associated hyper-parameters.

{\bf Contributions} \quad Our contributions are summarized as follows: 1)~we provide a dataset generation framework for causal inference in recommender systems, which may be of independent interest to other fields; 2)~we generate a semi-synthetic dataset, named CTAR, that enables the task of inferring the cause of user preference w.r.t.~tags; 3)~we present descriptive statistics regarding the CTAR dataset, along with three baseline methods; 4)~we also discuss several other research tasks that may benefit from using this dataset; 5)~finally, the CTAR dataset and generation codes have been released under the MIT License, available at \texttt{\url{https://github.com/KID-22/CTAR}}.

A preliminary version of the CTAR dataset has been used for user-tag preference prediction competition at the 2021 Pacific Causal Inference Conference{\footnote{\texttt{\url{https://competition.huaweicloud.com/information/1000041488/introduction}}}}. We have received more than 750 submissions from 116 participating teams. We hope that the CTAR dataset and its generation framework will facilitate both the research and development of causality based recommendation models.

\section{Preliminaries}
In this section, we introduce useful causality concepts and briefly review existing recommendation datasets.
\subsection{Causal Graphical Model and Missing Mechanism}

Causal graphical model \cite{pearl2009} is a powerful tool in causal inference and has also attracted much interest in recommender systems \cite{zhang2021causal}. A causal graphical model or causal graph is a graph consisting of variables (nodes) and directed edges. The direction of an edge indicates the causal direction, and the absence of an edge between two variables means that there is no direct causation between them. Many established tools such as \emph{d}-separation and \emph{do}-calculus \cite{pearl2009,aprimer} can help analyze the relationships among variables based on causal graph. In this work, we only consider directed acyclic graphs where there is no loop of edges, and we will use  ``node'' and ``variable'' interchangeably.
    
Recommender systems often face the  missing data problem. For example, the data of user and item features are often fully observed in online shopping systems, while the clicks, over all possible user-item pairs, are largely unobserved or missing. As stated in \cite{10.1093/biomet/63.3.581rubinmissing}, the mechanisms of missing data can be classified into three categories: missing completely at random (MCAR), missing at random (MAR) and missing not at random (MNAR). MCAR means that the missing mechanism is independent of data. The missingness depends only on observed data for MAR, while  it may also depend on unobserved factors with MNAR. 
All the three missing mechanisms will be included in our CTAR dataset through the use of missing graphs \cite{mohan2021,DBLP:conf/nips/MohanPT13} that will be elaborated in \Cref{mgraphmgraph}.

\subsection{Recommendation Models and Related Datasets}
\label{biasissue}
%%%%%%%%%%%%%%%%%%%%%%
% Dataset Comparison
%%%%%%%%%%%%%%%%%%%%%%
% \begin{table*}[htp]
% \centering
% \caption{Comparison of existing datasets for unbiased and causal recommendation (\Checkmark ~| \XSolidBrush ~means totally | partially or not met, respectively)}
% \begin{adjustbox}{max width=0.8\textwidth}
% \begin{tabular}{cccccc}
% \toprule
%          & Counterfactuals & Unconfoundedness & Causal Explanations & Unbiased Testing & Flexibility \\
% \midrule
% MovieLens & \XSolidBrush & \XSolidBrush                & \XSolidBrush                   & \XSolidBrush                & \XSolidBrush         \\
% Yahoo!R3 & \XSolidBrush & \XSolidBrush                & \XSolidBrush                   & \Checkmark                & \XSolidBrush         \\
% Coat     & \XSolidBrush & \XSolidBrush                & \XSolidBrush                   & \Checkmark                & \XSolidBrush         \\
% MSSD     & \XSolidBrush & \XSolidBrush                & \XSolidBrush                   & \Checkmark                & \XSolidBrush        \\
% CTAR     & \Checkmark   & \Checkmark                & \Checkmark                   & \Checkmark                & \Checkmark       \\
% \bottomrule
% \end{tabular}
% \end{adjustbox}
% \label{tab-dataset}
% \end{table*}
%%%%%%%%%%%%%%%%%%%%%%

\begin{table}[tb]
\centering
\small
\caption{Comparison of existing datasets for unbiased and causal recommendation (\Checkmark ~| \XSolidBrush ~means totally | partially or not met, respectively)}
\begin{adjustbox}{max width=0.47\textwidth}
\begin{tabular}{lccccc}
\toprule
         & MovieLens & Yahoo!R3 & Coat & MSSD & CTAR \\
\midrule
Counterfactuals & \XSolidBrush & \XSolidBrush                & \XSolidBrush                   & \XSolidBrush                & \Checkmark         \\
Unconfoundedness & \XSolidBrush & \XSolidBrush                & \XSolidBrush                   & \XSolidBrush                & \Checkmark         \\
Causal Explanations     & \XSolidBrush & \XSolidBrush                & \XSolidBrush                   & \XSolidBrush                & \Checkmark         \\
Unbiased Testing     & \XSolidBrush & \Checkmark                & \Checkmark                   & \Checkmark                & \Checkmark        \\
Flexibility     & \XSolidBrush   & \XSolidBrush                & \XSolidBrush                   & \XSolidBrush                & \Checkmark       \\
\bottomrule
\end{tabular}
\end{adjustbox}
\label{tab-dataset}
\end{table}

As the core component of a recommender system, recommendation models have been thoroughly studied in the past decades and various models are proposed, including collaborative filtering \cite{koren2009matrix,KDD_WangWY15}, factorization machines \cite{rendle2010factorization,SIGIR_RendleGFS11,RecSys_JuanZCL16}, and deep neural network based models \cite{cheng2016wide,guo2018deepfm,RecSys_HuangZZ19}. Most of these models focus on user-item rating prediction tasks, but often omit the underlying \emph{causes} behind users' ratings and lack reliable explanations. Besides, they are usually learned based on observed data and are skewed due to the closed feedback loop in recommender systems, resulting in \emph{``the rich get richer''} Matthew effect \cite{schnabel2016recommendations,wang2019doubly,dong2020counterfactual}. In particular, the observed data possibly contains many biases including position bias \cite{agarwal2019estimating, yuan2020unbiased}, item exposure bias \cite{schnabel2016recommendations,yuan2019improving}, user self-selection bias \cite{saito2020asymmetric,WSDM_xjwrzysjzq21}, and popularity bias \cite{zhang2021causal,abdollahpouri2017controlling}, which are likely to result in biased models and affect users’ experiences.

{There are several benchmarked datasets for unbiased recommendation, e.g., Yahoo!R3~\cite{marlin2009collaborative}, Coat~\cite{schnabel2016recommendations}, MSSD~\cite{brost2019music}, and MovieLens~\cite{MovieLens}, etc., which have been widely used to develop and evaluate recommendation models. 
Table~\ref{tab-dataset} presents a comparison of our proposed dataset with the existing datasets for unbiased and causal recommendation. 
Although Yahoo!R3, Coat and MSSD contain unbiased testing set for evaluation, these observational datasets can still limit the causal discovery, estimation, and evaluation in recommendation studies due to several limitations. 
The first problem is the lack of ground-truth, that is, we can never observe the counterfactual (the other potential outcome) of an observation. In contrast, all counterfactuals are known in the CTAR dataset. Besides, when developing recommendation methods in the causal graphical model framework, the causal graph is usually assumed to be known \emph{a priori}. However, the causal graph is in general not testable from observational datasets. Our proposed CTAR can guarantee the correctness and unconfoundedness of the causal graph, thus enabling more reliable evaluation of recommendation models. There are other limitations in terms of dataset size, possible selection bias in the testing set, causal interpretations, and flexibility. For example, the Coat dataset has only a limited number of data samples. Although Yahoo!R3 contains some uniform data, it may suffer from selection bias because only users with more than 10 ratings are considered. Additionally, Yahoo!R3 and Coat only have rating data, while we also provide the tag data that can be used to discover the causes behind users’ preferences; see details in \Cref{CTAR}. Last but not least, our proposed framework enables the generation of various versions of CTAR with different types and levels of biases, thus CTAR is much more flexible than existing datasets. Note that MovieLens is constructed from biased logged data, which does not support unbiased evaluation directly. It is usually used for semi-synthetic dataset generation in unbiased recommendation~\cite{schnabel2016recommendations,wang2019doubly}, but it also suffers from the confoundedness issue, the lack of causes of user preference on items, and the lack of flexibility.}

\begin{figure}[b]%靠文字内容的右侧
\centering
% \vspace{-0.4cm}
\setlength{\abovecaptionskip}{0.cm}
\includegraphics[scale=0.65]{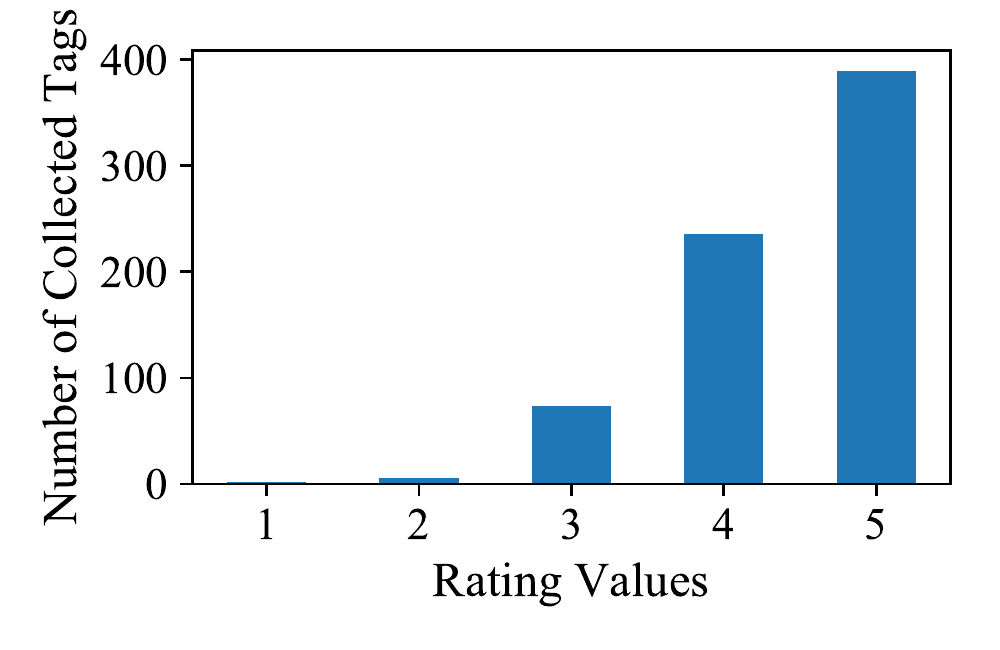}
\caption{Number of collected tags against ratings to  in the survey data.}
\label{collectags}
\end{figure}

{
\section{Problem formulation, Data Collection and Design}\label{datasetdesign}
In this section, we describe and give a formal definition to the problem of interest, and discuss the reason for the introduction of a new semi-synthetic dataset.
\subsection{A Need For Causal Tags}
In this work, we introduce the concept of ``causal tags'' as a way of explaining the reasons behind ratings. A causal tag means a particular feature of the movies that can affect user ratings and are represented by single words or phrases, e.g., director names or movie genres. The cause of a tag to rating can be defined  through causal effects using the \emph{do}-intervention \cite{pearl2009}, i.e., the difference between counterfactual ratings with and without this tag. If there is a difference in a movie rating with and without a particular tag, we say that the tag is one of the reasons why a user likes or dislikes the movie. For example, imagine different versions of the movie \textit{Sherlock Holmes}, where the directors or actors are different from the rest contents are kept similar in some sense. If there is any difference between the ratings of different versions, we can say that the directors or actors cause the changes in ratings and represent the user preference and we will give this quantity a formal definition in the next section. 
\par There are several potential needs for the introduction of causal tags in recommender systems. The first is that causal tags mean the real preference of users, which can support the development of personalized recommendation and a good personalized recommender system can promote the development of huge markets in many areas. The second need falls into the category of variable selection. Suppose $X_c\subset X$ where $X$ is the entire feature vector and $X_c$ is the subset of $X$ that has causal relationship with the outcome of interest, then the models trained on $\{Y,X_c\}$ may be better than those trained on $\{Y,X\}$. Consider the inverse propensity score model, where a propensity score $\pi(x)=P(T=1|X=x)$, meaning the probability of being assigned to a treatment, is estimated using relevant feature $X$ where $T=1$ denotes treatment assignment. In this case, the use of $X$ may lead to larger variance of the estimated $\hat{\pi}(x)$ \cite{hernan2020whatif} and harm the efficiency of the estimation. Thus the use of $X_c$ is preferred here.  
\subsection{Causal Tags And Causal Estimand of Interest}\label{sec_causaltagsandestimand}
When talking about causal inference, a typical approach is to first translate the scientific question under study into a well-defined causal estimand before adopting a model to estimate it \cite{wu2022causal}. Now we give a formal definition to the causal tags and discuss related issues with it. Using the \emph{do}-intervention terminology, the causal tag can be defined as follows. 
\begin{definition}[Causal Tag]
    A tag $T_i$ is called a causal tag of user $u$ if the causal estimand $\tau(u,T_i)$ defined below differs from 0 where $T_i=1$ represents having this tag in movies and $T_i=0$ otherwise.
    \begin{equation}\label{causaltag}
    % \begin{split}
        \tau(u,T_i)=\mathbb E[Rating|~u,~do(T_i=1)] - \mathbb E[Rating|~u,~do(T_i=0)].
    % \end{split}
\end{equation}
\end{definition}
It is obvious that $\tau(u,T_i)$ reflects a user’s preference on average and this quantity can be useful to a wide extent. The expectation in \Cref{causaltag} is taken over the entire (user, movie) pairs under consideration and in this sense, it treats the other tags $T_j$ for $j\ne i$ as features and taking expectation means taking average over these tags. One problem associated with \Cref{causaltag} is that the causal estimand defined in this way typically falls into the category of estimation problem or unsupervised learning problem, which means that although we can always make an estimation using the rating data, we can never verify whether our estimation is correct, and this is another reason why existing datasets are not suitable for inferring reasons, apart from those mentioned in \Cref{Introduction,biasissue}.
\par To deal with this unsupervised learning problem, a simpler version of causal estimand that defines a causal tag is given by
% If we are interested in a user's preference on a particular movie, we may consider the following quantity 
\begin{equation}\label{causaltagonmovie}
    \begin{split}
        \tau^{\prime}(u,T_i)=I\{&\mathbb E[Rating|~u,~do(T_i=1)] \\
        &- \mathbb E[Rating|~u,~do(T_i=0)]>0\},
    \end{split}
\end{equation}
where $I(\cdot)$ is the indicator function. Thus, while \Cref{causaltag} reflects how much a user likes a tag, \Cref{causaltagonmovie} reflects whether a user likes a tag. It follows immediately that the prediction of the latter one is a more simple question since we can somehow collect data on it and make it a supervised learning problem, while it's probably impossible to collect precise numbers on how much a user likes a movie, although the former one may be more helpful in aiding businesses like personalized recommendation. This fact reflects a trade-off in defining causal estimands and $\tau^{\prime}(u,T_i)$ is the motivating estimand of our task.
\par Note that except for the ones defined in \Cref{causaltag} and \Cref{causaltagonmovie}, other estimands are possible up to one's need. For example, we have treated other tags as features earlier, while it is possible to include more tags into the \emph{do}-operation, which will give an opportunity on studying the interaction causal effects between tags, etc. With the causal estimands defined, where each estimand corresponds to a certain problem of interest, we can then analyze whether we have appropriate data to answer these questions.
\subsection{Need for Synthetic Dataset}
}
To enable the task of inferring $\tau^{\prime}(u,T_i)$,
% of inferring the causes underlying users' preferences, 
we attempted to conduct an online survey where subjects were asked to rate and apply \emph{causal} tags to some selected movies at the very beginning. 
% Here the causal tags,  out of all the related tags of a movie,  are those that indicate the reasons why they like or dislike the movies. 
We designed the survey's user interface based on a cultural tagging study \cite{dong2012war} to minimize the textual content that might influence their choice, and selected a number of movies based on their popularity measured by the number of ratings in a popular movie review website. For sanity check, we included 50 popular movies and 273 distinct tags in the initial survey. The survey was carried on for two months and eventually we collected in total 707 ratings and 701 causal tags from 552 subjects. A summary of the number of causal tags versus ratings is further shown in \Cref{collectags}.

\par From this initial surveyed data, we first observed that user response is generally sparse---there are less than 1.3 ratings or tags from a subject on average. Moreover, as shown in \Cref{collectags}, the subjects tend to rate and apply causal tags to the movies they like. Indeed, the unconfoundedness assumption can hardly be satisfied in practice. There are also potential compliance issues, e.g., some users may provide arbitrary ratings and tags despite that they are assumed to be the underlying truth. Additionally,  selection bias is likely to exist due to  survey coverage. To be noted, the confounding effect, compliance issue, and selection bias are hard to avoid in the survey approach, even if more subjects are involved.
\par Due to the above limitations, we alternatively aim at semi-synthetic datasets with guaranteed causal interpretations. The previous surveyed data, nevertheless, guide us to design practically meaningful causal mechanisms in our approach. To begin with, we first collected observational data of movies with their descriptive tags and rating information from a famous movie rating website, resulting in a dataset consisting of 9,715 distinct movies, 10,273 distinct tags, and 75,460 ratings (the rating scale is from 1 to 5). The collected data will  be used to determine some hyper-parameters in our semi-synthetic recommendation dataset. For example,  the average rating of each movie in the collected data are used as the base rating and the total number of tags associated with a movie serves as the popularity criterion. A summary of these collected data are also released for verifying the generation procedure of the proposed framework.

\section{Causal Graphical Models with Missingness and Identifiability Issues}
This section introduces the causal graphical models with missingness for our dataset. We also discuss the identifiability issue in causal inference.

\subsection{Missingness Graph for Data Generation}
\label{mgraphmgraph}
In this work, we utilize missingness graphs or m-graphs \cite{mohan2021,DBLP:conf/nips/MohanPT13} to generate our CTAR datasets.  To the best of our knowledge, we are the first to introduce m-graphs into recommender systems, considering the pervasive missing data problem in real scenarios. Comparing with regular causal graphs, m-graphs provide an explicit and intuitive way of dealing with missing mechanisms.

We first present a formal definition of  m-graphs. Following the same terminology as in \cite{mohan2021}, we use $\mathbf{V}$ to denote the full set of nodes in an m-graph, which consists of five  subsets:
% Then it can be decomposed into 5 subsets:s
\begin{align}
\label{def:mgraph}
        \mathbf{V}=V_o\cup V_m\cup N \cup V^*\cup R,
\end{align}
with $V_o$ denoting the set of fully observed variables, $V_m$ the set of partially observed variables, $N$ the set of unobserved variables, $ V^*$ the set of proxy variables, and $R$ the set of indicators that represent the missing status of variables in $V_m$. For each partially observed variable $X\in V_m$, an m-graph defines two associated variables $X^*$ and $R_X$, where $X^*$ is the proxy variable actually observed and $R_X$ is the missing indicator. That is, $X^*=X\odot R_X$ where $R_X\in\{0,1\}^{|X|}$ and $|X|$ denotes the dimension of $X$. 

\par
\begin{figure}[tp]
          \centering
          \includegraphics[scale=0.8]{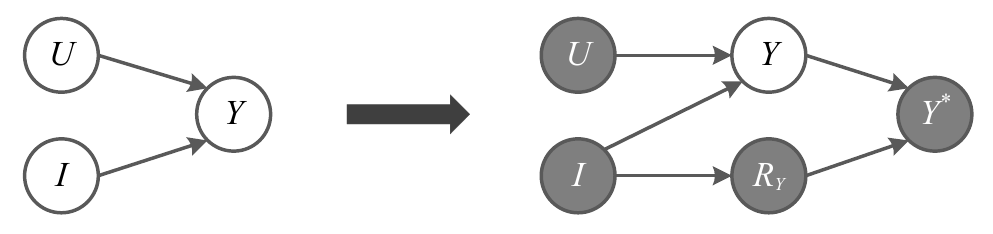}
          \caption{Traditional causal graph (left) and m-graph (right)}
          \label{tgraphvsmgraph}
        \end{figure}

 Causal graphical models describe the causal relationships among variables and also reflect  the data generation process. However, they may ignore the data collection process where some variables may be missing. An m-graph, on the other hand, extracts and delineates the missing mechanisms by  adding two new types of nodes in the graph, and  can better reveal the missing mechanisms in data collection. Here we use an example to illustrate the benefits of m-graphs in recommender systems. Shown in the left panel of  \Cref{tgraphvsmgraph} is a simple causal graph describing the relationships between user {$U$, item $I$}, and the outcome of interest $Y$ like rating or click.
% In \Cref{tgraphvsmgraph}, the  causal graph is on the left,
Suppose here that data are generated by a popularity based recommender system and the popularity bias is reflected by $ I\rightarrow Y$. Meanwhile, this causation may also be due to  that higher quality items tend to have higher ratings. In the corresponding m-graph in \Cref{tgraphvsmgraph}, $ Y^{*} $ is the observed outcome that serves as a proxy to the no-missing outcome  $Y$ and $R_Y$ stands for the  mechanism that causes the missingness in $ Y^*$. One can verify that  the popularity bias in the m-graph is now represented by $ I\rightarrow R_y\rightarrow Y^*$ while the effect of item quality is given by $ I\rightarrow Y $. Notice  that in this m-graph, we have $ V_o=\{U,I\} $, $ V_m=\{Y\} $, $ N=\emptyset $, $ V^*=\{Y^*\} $, and $ R=\{R_Y\} $, according to the definition in \Cref{def:mgraph}.
        %(may include other advantages like help to clarify assumptions and definitions of nodes and tables  illustrating missing mechanisms)
        
{\bf Remark.}\quad {Since the variables in $V^*$ are always the children of  variables in $R$ and $V_m$, we may omit $V^*$ in the m-graph if it is not involved in our analysis. For example, when determining the type of missingness, we are interested in  % only focus on 
the relationships among $V_o,V_m,N,R$ and thus $V^*$ can be ignored.
% (how to determine the type of missingness is discussed in \Cref{defmissing}). 
In what follows, we will state explicitly 
%in the remaining text when
if $V^*$ variables are suppressed. %omitt.
}

\subsection{Different Types of Missing Data}\label{defmissing}
 As first introduced in \citep{10.1093/biomet/63.3.581rubinmissing}, missing data problem has become an important issue in causal inference \cite{little2019statistical,zhou2014applied}, and is considered in many  recommender system literatures \cite{WSDM_xjwrzysjzq21,schnabel2016recommendations,saito2020asymmetric}. An advantage of an m-graph compared to a conventional causal graph is that, while it is fully compatible with conventional causal graph framework proposed in \citep{pearl2009}, it can handle the problem of missing data more clearly. Here we briefly introduce the concept of missing data under the m-graph framework \cite{mohan2021}. 
\par There are typically three types of missing data, namely MCAR (missing completely at random), MAR (missing at random) and MNAR (missing not at random). Based on the statistical dependencies between the missing mechanisms ($ R $) and the variables in the dataset ($ V_m,V_o $), \citet{mohan2021} gives a formal definition of these missingness in m-graph framework as follows:
    \begin{enumerate}
        \item Data are MCAR if $ V_m \cup V_o \cup N\indep R $ holds in the m-graph. In words, missingness occurs completely at random and is entirely independent of both the observed and the partially observed variables. This condition can be easily identified in an m-graph by the absence of edges between the $ R $ variables and variables in $ V_o \cup V_m $.
        \item Data are MAR if $ V_m\cup N\indep R|V_o $ holds in the m-graph. In words, conditional on the fully observed variables $ V_o $, missingness occurs at random. In graphical terms, MAR holds if (i) no edges exist between an $ R $ variable and any partially observed variable and (ii) no bidirected edge exists between an $ R $ variable and a fully observed variable. MCAR implies MAR, ergo all estimation techniques applicable to MAR can be safely applied to MCAR.
        \item Data that are not MAR or MCAR fall under the MNAR category.
    \end{enumerate}
As an example, consider the m-graph in \Cref{tgraphvsmgraph}. Since $Y\indep R_Y|I$ while $Y\nind R_Y$ due to the path $Y\leftarrow I\rightarrow R_Y$, we conclude that the missing mechanism behind $Y$ is MAR.
% \par {In our CTAR dataset, $R_\text{RCT}$ belongs to MCAR, $R_R$ is MAR and $R_O$ belongs to MNAR due to  the path $T_L\rightarrow R_O$ where $T_L\in V_m$ in \Cref{ourfullappendix}. Since $RecSys, T_U, Q$ are all mediators, they can by encoded into the the edges they connect to, omitted in other words, in the analysis of missing data without loss of accuracy. For example, we can simple replace $U\rightarrow T_U\rightarrow T_L$ with $U\rightarrow  T_L$ without changing the missing mechanism behind our m-graph.} The independence and conditional independence mentioned in the above definition can all be checked by the \textit{d}-separation tool introduced in \cite{pearl2009}.

\subsection{Identifiability Issues}\label{recover}

Identifiability is a critical issue in causal inference, but to our best knowledge it has been rarely discussed in recommender systems. {Indeed, it is a prerequisite in the causal inference based recommendation models and guaranteed estimation and evaluation rely on the identifiability.}
% \deng{A causal estimand defined by counterfactual variables is said not to be identifiable if there exist two different choices of causal estimands under which the observed data have the same distribution. Otherwise, the causal estimand is identifiable.}
Roughly speaking, identifiability in the context of causal inference means whether we can use observable quantities to estimate the unobservable counterfactual estimands with consistency. Identifiability is also called \textit{recoverability} in \cite{mohan2021}. {The significance of discussing identifiability is at least twofold: first, we can ascertain whether a consistent (or unbiased) estimate of the counterfactual estimand of interest can be obtained from the data available under some reasonable assumption; second, if the estimand is identifiable, we can explicitly present the identifiability assumptions underlying the estimation approaches. This provides a desirable perspective to evaluate the debiasing methods by assessing the assumptions and provides an opportunity to develop new approaches by weakening the assumptions.} A formal definition is given as follows.

\begin{table}[tp]
\small
\centering
\setlength{\abovecaptionskip}{0.1cm}
\caption{Description of the nodes and symbols used in our dataset}
\label{tab:datasetnames}
{\small
\begin{tabular}{|p{1.0cm}|p{0.75cm}|p{5.7cm}|} 
    \hline
                            Name      & Symbol & Description   \\ 
    \hline
    \hline
        Movie & $ M $ & Movie, a variable denoting movie ID \\
    \hline
         User & $ U $ & User, a variable denoting user ID \\
    \hline
    \multirow{2}{*}{TagMovie} & \multirow{2}{*}{$ T_M $} & Tags associated with a movie, such as ``Love'' for \textit{Flipped} \\
    \hline
    \multirow{2}{*}{TagUser} & \multirow{2}{*}{$ T_U $} & Tags that indicate user preference, e.g., ``Romantic'' for some users \\
    \hline
    TagLike & $ T_L $ & Overlapped tags of a movie and a user \\
    \hline
    Quality & $ Q $ & Movie intrinsic feature \\
    \hline
    Rating & $ R $ & Rating \\
    \hline
    %\ly{RCTTag & $-$ & A training set containing RCT samples drawn from $T_L$}\\
    %\ly{ObsTag & $-$ & A training set containing observational samples drawn from $T_L$}\\
    \multirow{2}{*}{$RCTTag^*$} & \multirow{2}{*}{$ RCT^* $} & Observed tags indicating a user's preference for a movie, collected through RCT experiment {from $T_L$} \\
    \hline
    \multirow{3}{*}{$ObsTag^*$} & \multirow{3}{*}{$ O^* $} & Observed tags of a movie that indicate a user's preference, collected through observational experiment {from $T_L$} \\
    \hline
    $Rating^*$ & $ R^* $ & Observed ratings \\
    \hline
    $R_\text{RCTTag}$ & $ R_\text{RCT} $ & Missing mechanism associated with $RCT^*$ \\
    \hline
    $R_\text{ObsTag}$ & $ R_\text{O} $ & Missing mechanism associated with $O^*$ \\
    \hline
    $R_\text{Rating}$ & $ R_\text{R} $ & Missing mechanism associated with $R^*$ \\
    \hline
\end{tabular}
}
\end{table}

\begin{definition}[Identifiability of target quantity \cite{mohan2021}]  
    Let $\mathcal{A}$ denote the set of assumptions on the data generation process and let $Q$ be any functional of the underlying distribution $P(V_m, V_o, R)$. Then $Q$ is said to be identifiable if there exists a procedure that can compute a consistent estimate of  $Q$ for all strictly positive distributions $P(V^*, V_o, R)$ w.r.t.~the observed data that are generated under $\mathcal{A}$.\label{recoverdef}
\end{definition}

    \begin{figure}[b]
      \centering
      \vspace{-0.3cm}
      \includegraphics[scale=0.37]{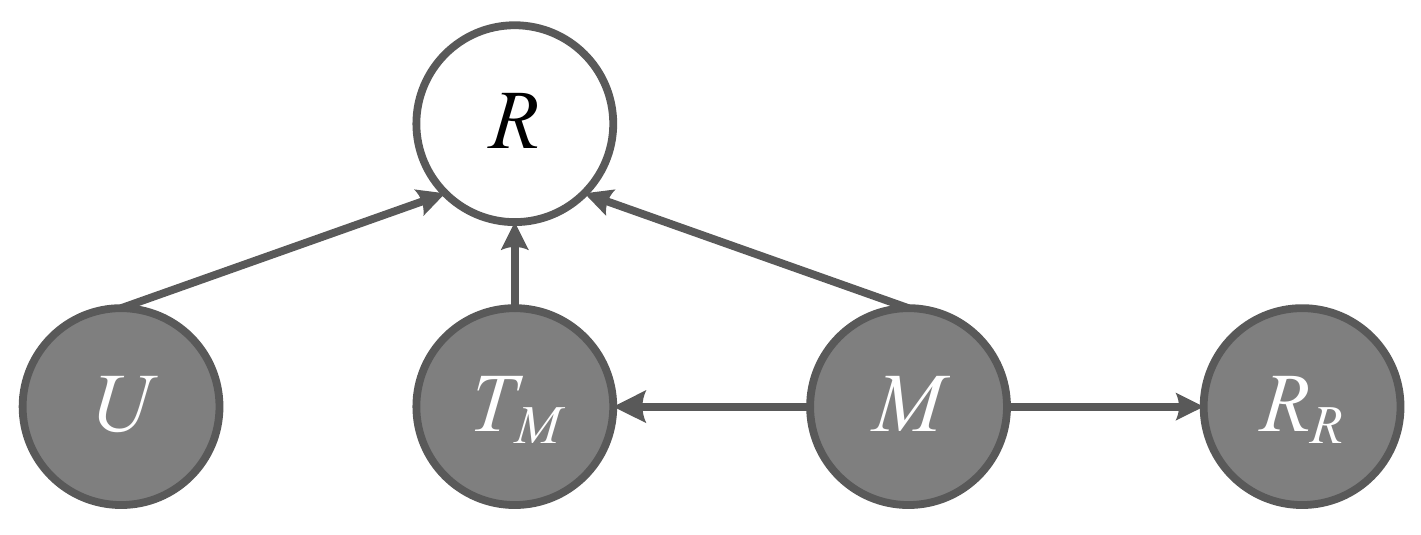}
      \caption{A simple model with missing mechanism. Meaning of notations:  $R$: rating, $U$: user, $M$: movie, $T_M$: the set of movie tags, $R_R$: the missing mechanism associated with $R$. Here we do not include $R^*$ for simplicity.}
      \label{oursimple}
    \end{figure}
    
Missingness due to biases in the data can have an impact on the identifiablity: while MCAR and MAR do not affect the identifiability of target quantity,   MNAR generally does except for some special cases \cite{mohan2021}. In real datasets, it may be difficult to know exactly the underlying missing mechanisms, and if MNAR exists, then the target quantity may become non-identifiable. Consequently,  evaluation of the recommendation methods w.r.t.~this target quantity may be questionable, as the quantity of interest is non-identifiable to any method. In contrast, in a semi-synthetic dataset, such as the proposed CTAR dataset in the next section, we have full control over both the target quantity and missing mechanisms. Therefore, we know exactly whether the interested quantity, which may be affected with missing mechanisms, is identifiable or not. Consider \Cref{oursimple}, which illustrates a simple model for generating ratings ($R$) from users ($U$) and movies ($M$) with missingness, as an example. The missing mechanism behind \Cref{oursimple} is MAR, as $R$ is independent of $R_R$ conditional on the rest three nodes, and thus identifiable. 
% {More details about the choice of missingness in an m-graph can be found in \Cref{defmissing}.}
% Note that we do not include $R^*$ in \Cref{oursimple} because we may omit proxy variables $V^*$ when analyzing missingness.

\section{Dataset Generation for Recommendation}
\label{DGRE}
{This section describes how the proposed framework can be used to generate the CTAR dataset, based on the collected movie data and m-graphs.}
% This section describes the generation of the CTAR dataset, based on the collected movie data and m-graphs. 

\Cref{tab:datasetnames} summarizes the notations used throughout this paper. {Here Movie and User represent the real movies and users in a recommender system. For a user, TagUser is a list of tags that represents his/her preferred tags. For a movie, TagMovie includes the tags that represents the feature or content of this movie. TagLike is the intersection of TagUser and TagMoive,  the overlap of a user's preference and a movie's feature. Quality of a movie may not be described by tags but do have effects on the rating. All the nodes described above can be treated as ground truth and do not have missing values. As described in \Cref{mgraphmgraph}, all nodes starting with $R$ (except Rating) are the missing mechanisms associated with the corresponding nodes,} 
%\zhushengyu{and RCTTag and ObsTag are all sample sets of TagLike, where RCTTag represents a RCT sample while ObsTag is an observational sample that may be biased. } 
{ where RCTTag and ObsTag stands for different sampling methods of tags .} Here RCT (randomized controlled trial) is a data collection method that is usually regarded as  the ``golden standard'' for evaluating causal effect, {while Obs (observational) stands for observational method that may contain biases}. In our case, we consider that  $RCT^*$ is unbiased and the tags reflect the underlying truth of user preference over movies, while $O^*$ is collected through observational experiment where users may only label some of their preference tags, and is biased and noisy. Notice that the actual \emph{observed} tags in both $RCT^*$ and $O^*$ can still be affected by some missing mechanisms. 

%and data collected this way can be assumed to be unbiased.% \ly{We first summarize the variables, as well as description and symbols for each variable in \Cref{tab:datasetnames}.}

% \begin{figure}[htp]
%   \centering
%   \setlength{\abovecaptionskip}{0.cm}
%   \subfigure[]{\label{ourformu}
%     \includegraphics[scale=0.37]{image/ourinitial.pdf}
%   }
%   \subfigure[]{\label{ourcompress}
%     \includegraphics[scale=0.37]{image/ourcompress.pdf}
%   }
%   \caption{Formulation graph ({left}) and compression graph ({right}) for the proposed dataset.}
%   \label{ourinitialandcompress}
% \end{figure}

\subsection{Data Generation Workflow }
\label{FEC}

% We will use  the following three-step workflow to ease our data generation process.

% {\bf Formulation}\quad   The first step is to construct a causal graph that reflects the underlying mechanisms of primary interest. One shall also define the estimand explicitly in this step, which is expected to be identifiable. \Cref{ourformu} shows the graph representing the primary goal in our case. Here we aim to estimate the causal effect of tag $T_L$ on rating $R$, to explain why a user applies his/her rating to the movie w.r.t.~tags. 

\begin{figure}[t]
      \centering
      \includegraphics[scale=0.4]{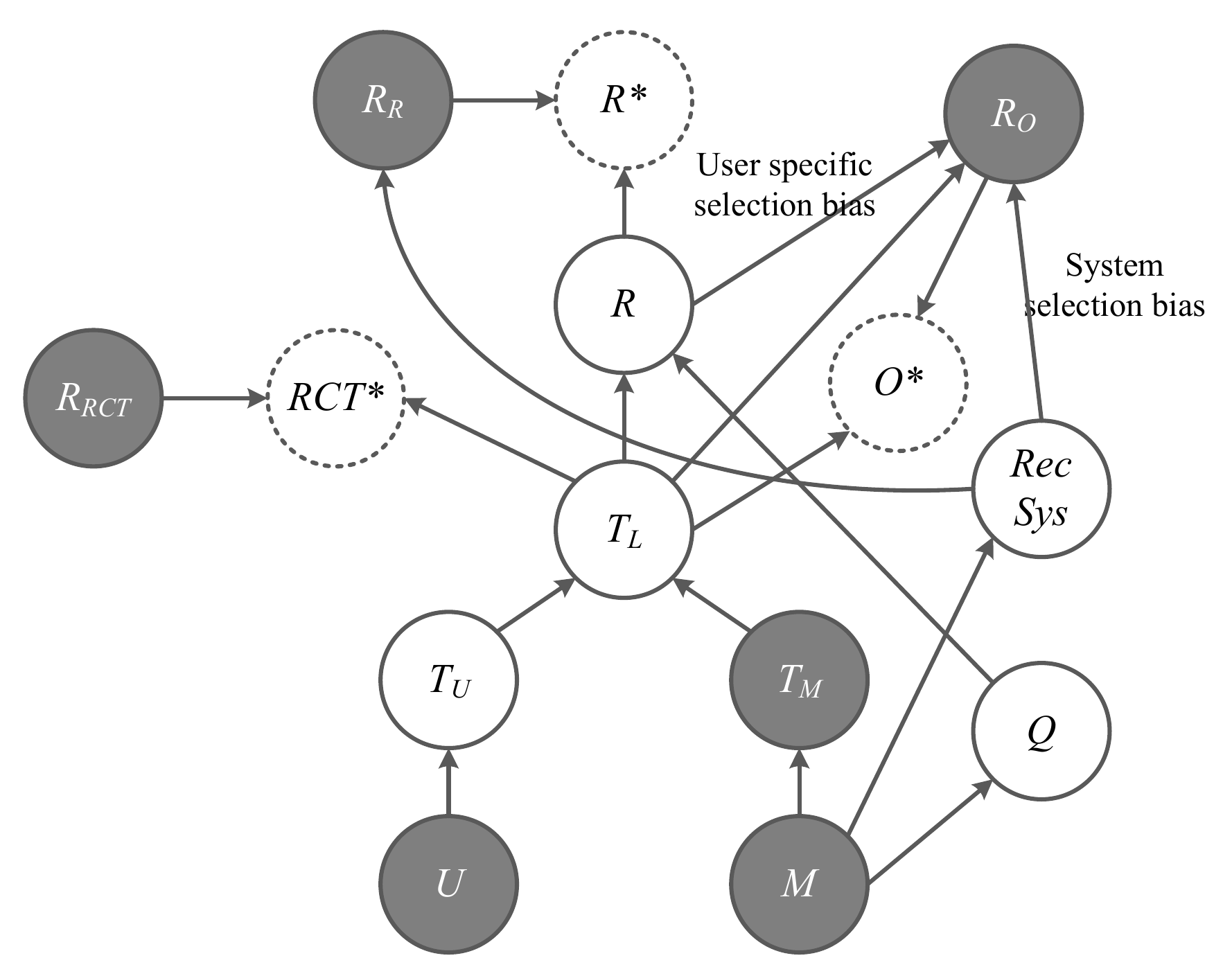}
      \caption{The m-graph for our semi-synthetic dataset. Definition of notations can be found in \Cref{tab:datasetnames}.} 
      \label{ourfull}
    \end{figure}

% {\bf Extraction} \quad    
To facilitate the data generation process, we need to construct an m-graph representing the data generation mechanism at first. Here we aim to estimate the causal effect of tag $T_L$ on rating $R$ in \Cref{ourfull}, to explain why a user applies his/her rating to the movie w.r.t.~tags. In the recommendation dataset, we consider to introduce the missing mechanisms for both rating $R$ and the additional observational information of $T_L$, resulting in the observed data $RCT^*$ and $O^*$. This setting can be treated as a data fusion problem. More details will be discussed in \Cref{ratingmissingsection,datafusionsection}, respectively. The complete data generating mechanism is given in \Cref{ourfull}, and a detailed description will be provided along with the released codes. 
    
    \par Different from \cite{mohan2021}, we define three types of nodes---black nodes, white nodes, and dashed nodes---for a better illustration. Here black nodes stand for fully observed data, including  both the fully observed variables and the associated missing mechanisms. In particular, black nodes in \Cref{ourfull} include the fully observed variables $U$, $M$ and $T_M$, and the missing mechanisms $R_R$, $R_\text{RCT}$ and $R_O$. White nodes represent all the unobserved variables, i.e., $T_U$, $T_L$, $Q$, recommender systems ($RecSys$) and $R$ in our case. Dashed nodes denote all the proxy variables, including $RCT^*$, $O^*$ and $R^*$. Please see the definitions of these notations  in \Cref{tab:datasetnames}. 
    
    \par Also note that in \Cref{ourfull}, $R_\text{RCT}$ belongs to MCAR, $R_R$ is MAR and $R_O$ belongs to MNAR due to  the path $T_L\rightarrow R_O$ where $T_L\in V_m$ in \Cref{ourfull}. 
    % Since $RecSys, T_U, Q$ are all mediators, they can by encoded into the the edges they connect to, omitted in other words, in the analysis of missing data without loss of accuracy. For example, we can simple replace $U\rightarrow T_U\rightarrow T_L$ with $U\rightarrow  T_L$ without changing the missing mechanism behind our m-graph.} 
    The independence and conditional independence mentioned in the above definition can all be checked by the \textit{d}-separation tool introduced in \cite{pearl2009}.
    
% {\bf Compression}\quad 
%   This final step serves as a sanity check of the proposed m-graph. A compression graph can be constructed by removing the nodes of the missing mechanism and proxy variables, and then dealing with the edges accordingly (details can be found in \Cref{excompression}). The corresponding compression graph in our graph is shown in \Cref{ourcompress}. Notice that when dealing with the edges in the new graph, we may use different types of edges to represent different types of causation. For example, in \Cref{ourcompress}, $M\rightarrow R^*$ means the relationship $R=f(M)$ and $M\dashrightarrow R^*$ represents popularity bias. We can then look into this new causal graph to check if it satisfies our initial purpose. 

        %Here we discuss this mechanism briefly and the detailed implementation of this mechanism is discussed in \Cref{datagen}.
    % \subsection{Different Types of Nodes}

\subsection{Rating and Its Missing Mechanism}\label{ratingmissingsection}
Based on the survey data, we assume that rating only depends on two factors: one is the interaction between movie and user preference, and the other is the movie's intrinsic feature like the quality. The former induces the heterogeneity of ratings among users and is represented by the path $T_L\rightarrow R$,  with $T_L$ being the the set of a movie's tags that are liked by the user. The latter  makes movie act as a confounder or common cause for  movie tags $T_M$ and rating $R$. This factor is realized by introducing movie intrinsic feature or quality $Q$ and a path $M\rightarrow Q \rightarrow R$. In our dataset, the value of $Q$ of a movie is generated based on the average rating of that movie in the  collected movie data. We use the following structural equations to generate the values of quality and rating:
        \begin{align}
        \label{quality}
            Q_m&=\mathcal{R}_m+\epsilon,\quad \epsilon\sim \mathcal {N}\left(0,\sigma_1^2\right),\\
            \label{rating}
            R_{u,m}&=Q_m+\epsilon,\quad \epsilon\sim \mathcal{N}\left(\dfrac{|T_L(u,m)|}{2}-\mu,\sigma_2^2\right)
        \end{align}
        where $u$ denotes a user, $ m $ denotes a movie, $|\cdot|$ is the cardinality of  a set,  $\mathcal{R}_m$ denotes the average rating obtained from the collected movie data, and $\mu$, $\sigma_1^2$, $\sigma_2^2$ are the parameters of noise variable $\epsilon$. 
        
In practice, the missing mechanism for rating may suffer from various types of biases. Here we only include popularity bias that is reflected by the path  $RecSys \rightarrow R_R $ in \Cref{ourfull}. The missing rate of a  rating is determined according to
        \begin{equation}
            P_{\textrm{missing}}=c\times P_M\times  \operatorname{Sigmoid}\left(\frac{|M|-rank_m}{T}+b\right)\label{misrating}
        \end{equation}
where $b$ and $T$ are respectively the bias and temperature in the logistic-sigmoid function, $P_M$ represents the  missing rate, $rank_m$ stands for the rank or order of movie $m$ among all movies, $c$ is a normalization constant so that the average missing rate w.r.t.~$P_{\textrm{missing}}$ is equal to $P_M$. The  values of these parameters used in our CTAR dataset will be released in {our github repository}. \ly{It is also easy to verify that the type of missingness of $R$ is MAR, since conditional on $M$, $R_R\indep R$ in \Cref{ourfull}.}
        % randomized controlled trial
\subsection{Data Fusion for Observed Movie-User Tags}\label{datafusionsection}
%\zhushengyu{RCT has been defiend?}

%we are interested 
Recall that a primary task in this paper is to infer  the  cause  of  user  preference  through  estimating the causal effect of each tag.  \ly{As we have discussed in \Cref{sec_causaltagsandestimand}, this preference may be inferred using only the rating data, but hard to estimate and verify.}
% or we are interested in $\mathbb E[Rating|~user~u,~do(with~tag~A)] - \mathbb E[Rating|~user~u,~do(without~tag~A)]$. If the difference is positive, then we may conclude that the tag is a preference tag of the user. 
Thus, besides the rating data of movie-user pairs, in practice we may also \ly{seek a simpler causal estimand as defined in \Cref{causaltagonmovie} and }collect such preference tags of movie-user pairs by, e.g., conducting surveys as described in \Cref{datasetdesign}.  In our setting, which can be seen as a simplification of real scenario, such datasets are classified into the observed sets $RCT^*$ and $O^*$. $RCT^*$ contains RCT selections of the overlapped tags $T_L$, where the subjects are required to label \emph{all} the preference tags for some movies. And $O^*$ is collected through observational experiment where users may only label some of their preference tags and may be biased due to user selection bias, system selection bias, etc. \ly{The missingness for $RCT^*$ and $O^*$ are MCAR and MNAR respectively. This is because $RCT^*\indep R_{RCT}$ and $R_O\nind O^*|M$ due to the confounding effect of $T_L$.}

%This setting is to reflect the pratical setting where we may co

%\ly{ Recall that at first, we collect two datasets as mentioned in \Cref{datasetdesign}, and the introduce of $RCT^*$ and $O^*$ is to reflect this setting.}
        
The reason of introducing these two sets is that RCT data are desired but maybe expensive to obtain in practice. Consequently, there is a need to include observational data that tend to be biased. This setting is to reflect real scenarios and is called data fusion, an emerging topic in both recommender systems \cite{bonner2018causal, liu2020a} and many other fields \cite{DBLP:journals/pnas/BareinboimP16pearldatafusion,NEURIPS2018_566f0ea4kallusdatafusion}. In our case, the missing mechanism $R_\text{RCT}$ is  generated by randomly picking a user-movie pair and the corresponding overlapped tag set $T_L$. For $R_O$, we adopt a two-step selecting procedure in which we first select a user-movie pair and then randomly sample some tags from the corresponding set $T_L$. In the first step, the missing rate of a user-movie pair is given by
        \begin{equation}
            P_{\textrm{missing}}=c\times P_M\times  \operatorname{Sigmoid}\left(\frac{|M|-rank_m+\alpha R_{u,m}}{T}+b\right)\label{mistag}
        \end{equation}
where $ b,c, T,rank_m,P_M $ have the same meanings as defined in \Cref{misrating} and $\alpha$ is the weight of rating $R_{u,m}$ that determines the magnitude of user selection bias. If  $T_L$ for a user-movie pair has more than one tag, then we first randomly pick a tag and toss a coin for each of the remaining tags to decide whether it is missing. This procedure is represented by $T_L\rightarrow R_O$ and reflects the cases where a user may only  label some but not all the tags indicating his/her preferences in observational experiments.

\begin{figure*}[t]
  \centering
  \setlength{\abovecaptionskip}{0.cm}
  \includegraphics[scale=0.55]{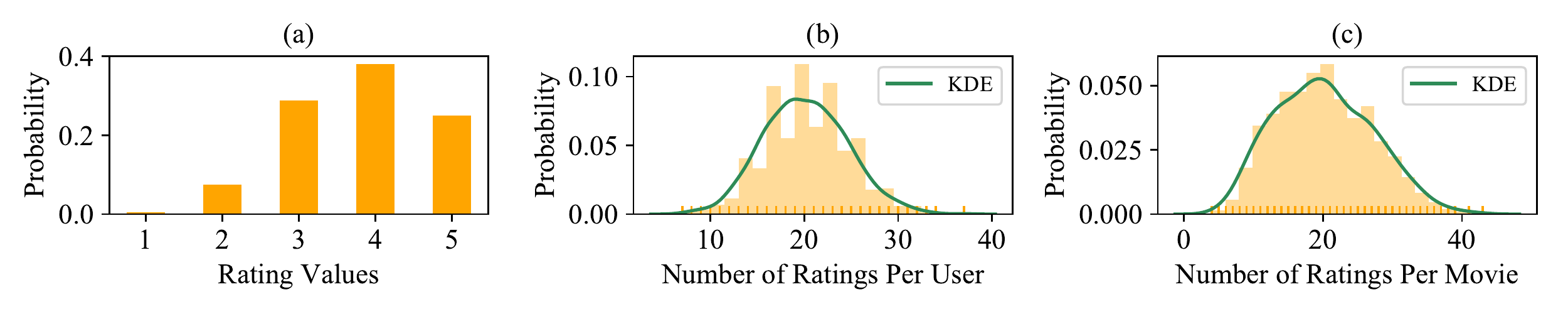}
  \caption{Distributions of rating set.}
  \label{summary_statistic}
  \vspace{-0.2cm}
\end{figure*}
\section{Causal Recommendation Dataset and Baseline Results}
\label{CTAR}

As stated earlier, instead of predicting user-movie ratings, we aim to infer the causes of user preference to a movie in terms of tags, based on the Causal Tag And Rating (CTAR) dataset that is generated following the proposed data generation framework. Knowing such information can also help explain the recommendation results. In this section, we present descriptive  statistics about the generated CTAR dataset, along with baseline results for the user-tag preference prediction task.

\subsection{Dataset Description }\label{datasetdeswithsetpartition}
The CTAR dataset describes the interactive behaviors of 1,000 users onto 1,000 movies, with user-movie ratings and user-movie tags. Here the movies are selected according to their popularity in the collected movie data while the users are randomly selected. We assume that each movie has its own descriptive tags (e.g., movie genre, director, actors, etc.) that are observed. Each user has his/her own preferences to certain tags which determine whether he/she likes or dislikes the movie. Note that these preference tags associated with users can be observed but are not necessarily complete and accurate.  The dataset can then be generated according to the topological order indicated by the causal graph and the described causal relationships in 
\Cref{DGRE}. %More details are given \ly{in our github repository}.

%This dataset contains data from a random experiment. 

CTAR has four sub-datasets that can be used for training: {\tt Movie}, {\tt Rating}, \texttt{ObsTag} and \texttt{RCTTag}. \texttt{Movie} consists of all the movies and their descriptive tags. \texttt{Rating} contains the observed ratings of some user-movie pairs and may suffer from the popularity bias existing in real recommender systems. \texttt{ObsTag} and \texttt{RCTTag} contain the observed tags that are labelled by users and indicate user preference to a number of movies. The difference is that data in \texttt{RCTTag} reflect all the underlying tags for a user-movie pair, while users may only label part of the preference tags in   \texttt{ObsTag}. A more detailed description is provided in \Cref{DatasetDescription}.

We divide the total Test Dataset into three sub-datasets. Dataset~I consist of all the missing values in $ O^* $ caused by $T_L\rightarrow R_O$. The user-tag data in Datasets~II and III come from user-movie-tag tuples with and without observed ratings, respectively. These test datatsets reflect the increasing difficulties for the inference task, for example, Dataset~II contains more information, i.e.,  ratings, than  Dataset~III. This is also verified by our baseline results in \Cref{baselineresult}. As such, we suggest to also report the evaluation results w.r.t.~each of the test dataset when using the CTAR datasets.

\subsection{Dataset Statisics}
\begin{figure}[htp]%靠文字内容的右侧
\centering
\setlength{\abovecaptionskip}{0.1 cm}
\includegraphics[scale=0.7]{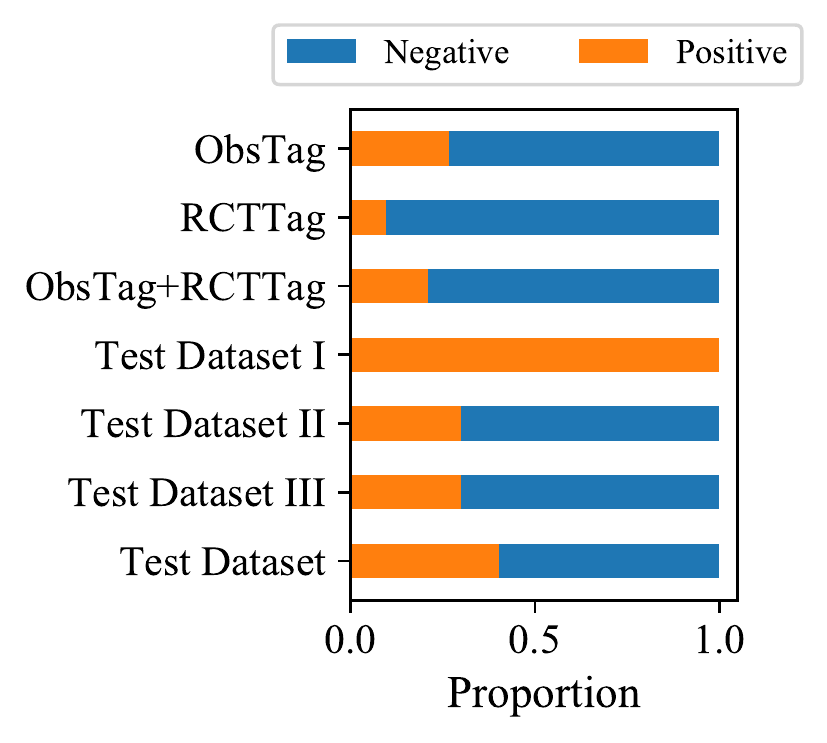}
\caption{Label distributions.}
\label{pos_neg_count}
\end{figure}

Figure~\ref{summary_statistic}~(a) reports the distribution of  ratings in the training dataset. Figures \ref{summary_statistic}~(b) and (c) describe the distribution of numbers of ratings w.r.t.~user and movie, respectively. We also show the proportions of the numbers of positive  and negative samples in each dataset in Figure~\ref{pos_neg_count}; here positive and negative samples correspond to the user-tag pairs with ``dislike'' and ``like'' labels, respectively. We can see that the proportion of positive samples in \texttt{RCTTag} is significantly lower than that of \texttt{ObsTag}. This is because \texttt{RCTTag} does not suffer from selection bias, so it is more likely to contain movies that users dislike. Notice that Test Dataset I consists of all positive samples as it includes  all the missing values in \texttt{ObsTag}.

\subsection{Baseline Results for the User-Tag Preference  Inference}\label{baselineresult}
Our task is to infer whether a user likes  or dislikes a given tag, and   can be  treated as a binary rating prediction problem. Thus, we choose three effective methods originally developed for the rating prediction task as our baselines:
 
 \begin{table*}[htbp]
  \small
  \setlength{\abovecaptionskip}{0.2 cm}
  \caption{Empirical results for the  user-tag preference inference task, averaged over 50 runs.}
  \label{tab:result}
  \resizebox{\textwidth}{!}{
  \begin{tabular}{cccccccc}
    \toprule
    & \multicolumn{1}{c}{Test Dataset~{I}} & \multicolumn{2}{c}{Test Dataset~\textrm{II}} & \multicolumn{2}{c}{Test Dataset~{III}} & \multicolumn{2}{c}{Test Dataset}\\
    \cmidrule(lr){2-2} \cmidrule(lr){3-4} \cmidrule(lr){5-6} \cmidrule(lr){7-8}
    Model & MSE & MSE & AUC & MSE & AUC & MSE & AUC\\
    \midrule
   \texttt{MF} & 0.6988 (0.0079)  & 0.2335 (0.0017) & \textbf{0.8560 (0.0010)} & 0.2796 (0.0002) & \textbf{0.8008 (0.0019)} & 0.3186 (0.0019) & \textbf{0.8609 (0.0008)}\\
    \texttt{MF-IPS} & \textbf{0.1257 (0.0044)} & \textbf{0.1642 (0.0016)} & 0.7941 (0.0016) & \textbf{0.2176 (0.0008)} & 0.6484 (0.0032) & \textbf{0.1789 (0.0005)} & 0.7987 (0.0013) \\
    \texttt{CausE} & 0.5643 (0.0124) & 0.2045 (0.0029) & 0.8553 (0.0007) & 0.2666 (0.0003) & 0.7996 (0.0020) & 0.2804 (0.0032) & 0.8603 (0.0007) \\
    \bottomrule
  \end{tabular}}

\end{table*}

\begin{itemize}[leftmargin=*]
\item \texttt{MF} \cite{koren2009matrix}  is a standard matrix factorization method that optimizes its parameters by minimizing mean squared  error (MSE) with some regularization terms.
\item \texttt{MF-IPS} \cite{schnabel2016recommendations} is based on the \texttt{MF} model and further uses the inverse propensity score (IPS) estimator for  unbiased evaluation. It requires only a small number of unbiased data  when estimating the propensity score with Naive Bayes.
\item \texttt{CausE} \cite{bonner2018causal} is a domain adaptation based
method and also relies on the MF model. It combines a large number of biased  but a small number of unbiased data for an improved prediction performance.
\end{itemize}

We provide training and implementation details in our open repository \footnote{\texttt{\url{https://github.com/KID-22/CTAR}}}.  Table \ref{tab:result} reports the inference performances in terms of MSE and the area under ROC curve (AUC) on the CTAR test datasets, with each reported value averaged over 50 runs. Here we do not  report the AUC for Dataset~I as it consists of all positive samples. Interestingly, \texttt{MF} outperforms \texttt{MF-IPS} and \texttt{CausE} in terms of AUC. \texttt{MF-IPS} achieves the best MSE but worst AUC, especially on Dataset~III. This shows that \texttt{MF-IPS} tend to overfit on the training dataset. Finally, we note that the performances of all three baseline methods have a worse performance on Dataset~{III} than on Dataset~\textrm{II}.

\section{Future Research Tasks}\label{futuretask}

Besides the tasks of predicting the rating of a user-movie pair or inferring user preference w.r.t.~tags, the released CTAR dataset allows various other research directions, some of which are listed below:

\begin{itemize}[leftmargin=*]
\item\ly{{\bf Combining different sources of data} \quad As discussed in \Cref{sec_causaltagsandestimand} and \Cref{datafusionsection}, the causal estimand $\tau(u,T_i)$ defined in \Cref{causaltag} contains more information, uses only the rating data but is hard to estimate and verify, while $\tau^{\prime}(u,T_i)$ defined in \Cref{causaltagonmovie} contains less information, needs the collection of $O$ and $RCT$ in our case, but is easy to get. It follows immediately that there is an opportunity to improve our recommender system if we can use $R$ to enhance the learning from $O$ and $RCT$ and vice versa.}
\item {\bf Debiased learning in recommender systems} \quad The CTAR dataset is generated using causal graphical model that simulates common biases and missingn mechanisms to make it as close to practical scenarios as possible.  All the counterfactuals in CTAR are available and can be used for evaluating novel causality based debiasing methods.
\item {\bf Explainable recommendations}\quad Most recommender systems aim to rank movies in a descending order according to predicted ratings. {In modern recommender systems, there is also a need to explain the predicted ratings and to improve personalized recommendation. The  task of inferring the cause of user preference to a movie is a rough way of explaining the recommendation results, and a further step can be taken by estimating the observable and counterfactual ratings in \Cref{causaltag}. The most difficult part in estimating \Cref{causaltag} is that we can not observe the counterfactual ratings in real world. If we can develop novel algorithms that estimate the counterfactual ratings accurately, we can then calculate the causal effects of interest, such as those defined in \Cref{causaltag} and \Cref{causaltagonmovie}, and further achieve more explainable recommendations. The proposed framework can provide both user-movie ratings and user-movie tags, and hence is suitable for this task.}
%The proposed CTAR dataset provides both user-movie ratings and user-movie tags, and  is applicable for explainable recommendation research. Indeed, the considered task of inferring the cause of user preference to a movie is a way of explaining the recommendation results.

\item {\bf  Counterfactual evaluation} \quad Traditional development and iteration of recommender models rely on large-scale online A/B tests, which are generally expensive,  time-consuming, and even unethical in some cases \cite{brost2019music, inbook}. Counterfactual evaluation has recently become a promising alternative  as it allows offline evaluations of the online metrics, leading to a substantial increase in experimentation agility. In our semi-synthetic CTAR  dataset, both factual and counterfactual outcomes are known, allowing it to serve as a benchmark for counterfactual evaluation methods. 
\end{itemize}

\section{Ethics}\label{ethics}
The trade-off between accuracy and privacy is a long-standing question for recommender systems. Previous studies have tried to enhance privacy and preserve accuracy of the information system with several approaches, such as distributed learning \cite{berkovsky2007enhancing, polat2008privacy}, differential privacy \cite{shen2014privacy}, and federated learning \cite{yang2019federated,chen2018federated}.
Causal inference can help  understand data and decision making mechanism, and has the potential to learn more about the users and further make better decision with fewer data. It is interesting to study how to  apply the above approaches \cite{berkovsky2007enhancing, polat2008privacy,shen2014privacy,yang2019federated,chen2018federated} or explore new technologies to handle the privacy issues in causal inference of recommender systems.

Meanwhile, we believe that a way to protect users' privacy is to know users better and to develop a powerful recommender system. When knowing little about users, someone may try to gather as much information as he/she can get from users and pack this information into a deep neural network, with a high probability of violating privacy. But if we can know users better, this situation can be avoided. For example, if we know the true causal mechanism behind users' preferences, we can provide the same recommendation with less information under users' authorization, which can certainly reduce the risk of leaking personal information.% Even if someone tries to deduce about individuals privacy, we are able to make up for it because we can better know the fault from the true causal mechanism.}

\section{Concluding Remarks}
In this paper, we have proposed a semi-synthetic data generation framework and constructed the CTAR  dataset for causal inference and  explanations in recommender systems. CTAR is automatically generated based on a collected movie dataset and the causal graphical model with missingness.  It enables the tasks of 
inferring causes behind user's ratings on the movies w.r.t.~tags, and can also be used for debiased learning and counterfactual evaluation. Descriptive statistics and baseline results regarding the  dataset are also reported. 

A potential limitation of the CTAR dataset is its synthetic nature. Only popularity bias and user selection bias are considered for now.  Nevertheless, our dataset generation framework allows to easily add additional nodes in the m-graph to include more kinds of biases. We also provide APIs to ease the generation of customized datasets, e.g., to include more movies and users. {An upper API will also be provided to extend the current m-graph to account for more biases.} In addition, the introduced data generation framework can be of independent interest to other applications. For example, telecommunication networks usually have a number of parameters  affecting the performances like throughput. However, obtaining a reliable evaluation of a set of parameters may require one or two weeks in practice, and it becomes time-consuming to find the optimal parameters \cite{ChuaiCLGWLZS19}. The proposed data generation framework can therefore be used, together with some historical data, to support the policy evaluation.

\appendix
\input{appendix}
\newpage
\bibliographystyle{ACM-Reference-Format}
\bibliography{sample-base}


\end{document}

%% file: appendix.tex
\section*{Appendix}

%\ly{This supplementary document is organized as follows: 1) section \ref{DatasetDescription}: a detailed description of our CTAR dataset; 2) section \ref{datagen}: the detailed generation process and summaray statistics of our CTAR dataset; 3) section \ref{Baselinesetup}: the setups of the benchmarks we use; 4) section \ref{defmissing}: a discussion of missingness in m-graph; 5) section \ref{excompression}: the compression step in our generation with its application.}
\begin{table}[htbp]
\centering
\small
\setlength{\abovecaptionskip}{0.1 cm}
\caption{Summary of CTAR dataset.}
\label{tab:datasetsummary}
\begin{tabular}{lllll} 
    \toprule
                           & Filename & Size & Records & Data in each record~   \\ 
    \midrule
    \multirow{4}{*}{train} & movie.csv & 34KB & 1,000 & movieID, list of associated tags       \\
                           & rating.csv & 190KB & 19,897 & userID, movieID, rating  \\
                           & obstag.csv & 99KB & 9,619 & userID, movieID, tagID   \\
                           & rcttag.csv & 16KB & 1,489 & userID, movieID, tagID   \\ 
    \midrule
    % \multirow{4}{*}{validation}  & valid\_1.csv  & 4KB & 458 & userid,movieid,islike  \\
    %                   & valid\_2.csv  & 14KB & 1,500 & userid,movieid,islike  \\
    %                   & valid\_3.csv  & 12KB & 1,200 & userid,movieid,islike  \\
    %                   & validation.csv  & 30KB & 3,158 & userid,movieid,islike  \\
    % \midrule
    \multirow{4}{*}{test}  & test\_1.csv  & 11KB & 1,267 & userID, tagID, islike  \\
                           & test\_2.csv  & 39KB & 4,170 & userID, tagID, islike  \\
                           & test\_3.csv  & 36KB & 3,719 & userID, tagID, islike  \\
                           & test.csv  & 86KB & 9,156 & userID, tagID, islike  \\
    \bottomrule
\end{tabular}
\end{table}

% \begin{figure*}[htp]
%       \centering
%       \setlength{\abovecaptionskip}{0.cm}
%       \includegraphics[scale=0.45]{image/fullplusstepone.pdf}
%       \caption{M-graph of our proposed synthetic dataset (left) and the resulting graph for our first step of compression (right)}
%       \label{ourfullappendix}
% \end{figure*}

\section{Detailed Dataset Description}\label{DatasetDescription}
As summarized in  \Cref{tab:datasetsummary}, CTAR has four sub-datasets for training: {\tt Movie}, {\tt Rating}, \texttt{ObsTag} and \texttt{RCTTag}.
\begin{itemize}
\item\texttt{Movie}:
This dataset gives all the movies and their associated tags. We assume that each movie has only 8 distinct tags.
\item \texttt{Rating}:
This dataset contains the observed ratings of some user-movie pairs and it may suffer from the common biases existing in real world data. Rating of a particular user-movie pair mainly comes from two factors: movie intrinsic feature and user's preference to that movie. The former can be treated as heterogeneity among movies. For the latter, we assume that a user's preference of  a movie only depends on the number of movie tags the user likes. That is, users tend to give a  higher rate if a movie contains more tags that he/she likes. 
\item \texttt{ObsTag}:
This dataset contains the observed tags that are labelled by users for movies. We assume that users only label the tags that they like. If , users do not like any tag of the a movie, then If the field ``tagID'' would be labeled as ``-1''. Notice that a user may label fewer tags than what he/she really likes. For example, given a user-movie pair, if the user labels ``love'', it means that ``love'' is one of the 8 tags associated with that movie and the user indeed likes the tag ``love''. In the meanwhile, the user may also like the other 7 tags and he/she may simply forget to label those tags by chance. Note that we have used tagID (e.g.,``1'') to replace the tags (e.g., ``love'').
\item \texttt{RCTTag}:
This dataset contains data from a random experiment. The users and the movies labeled by users are randomly selected. We assume that users are ``forced'' to label all  the tags they like for the movie. That is, if ``-1'' appears, then it means that the user does not like any tag of that movie. Besides, we assume that users only label the tags they indeed like in this dataset.
\end{itemize}
Note that our semi-synthetic data generation
framework does not limit the scale of users and movies. The number of users in the generated CTAR dataset can be as large as you want. 
% However, the scale of movies are limited to the number of movies (i.e., $9715$ distinct movies) in collected observational data because some basic information of movies in the collected data is used in generation process. 
Meanwhile, although we use some background information in the generation of movies' data, this is not a must in general since we can always give some background information. To better meet the real world big-data scenarios, a larger scale version of dataset with $100K$ users of CTAR will be released in our open repository soon.